\documentclass[conference]{IEEEtran}
\IEEEoverridecommandlockouts
\usepackage{cite}
\usepackage{amsmath,amssymb,amsfonts}
\usepackage{algorithmic}
\usepackage{graphicx}
\usepackage{textcomp}
\usepackage{xcolor}
\usepackage{multirow}  % Required for merging rows
\def\BibTeX{{\rm B\kern-.05em{\sc i\kern-.025em b}\kern-.08em
    T\kern-.1667em\lower.7ex\hbox{E}\kern-.125emX}}
\begin{document}

\title{Simultaneous Multi-Beam Radar with Full Range Resolution exploiting Space-Code Beamforming\\
% {\footnotesize \textsuperscript{*}}
\thanks{Note: This work has been submitted to the IEEE for possible publication. Copyright may be transferred without notice, after which this version may no longer be accessible}
}

\author{
\IEEEauthorblockN{P. Pon Surendar, \quad K. Giridhar}
\IEEEauthorblockA{\textit{TelWiSe Group, Department of Electrical Engineering} \\
\textit{Indian Institute of Technology Madras, Chennai 600036, INDIA}\\
% Chennai, India \\
\{surendar, k.giridhar\}@telwise-research.com}
% \and
% \IEEEauthorblockN{K. Giridhar}
% \IEEEauthorblockA{\textit{Department of Electrical Engineering}\\
% \textit{Indian Institute of Technology Madras, Chennai, India}\\
% Chennai, India \\
% k.giridhar@telwise-research.com}
% \and
% \and
% \and
% \and
}

\maketitle

\begin{abstract}
Conventional monostatic radar systems typically exhibit a trade-off between long-range target detection achieved through narrow beams and short-range wide-area surveillance employing broad beams. Realizing both functionalities within a single system, enabling simultaneous surveillance and long-range target localization, poses a significant challenge. This paper presents a novel signal model and an all-digital frequency domain radar architecture leveraging first-of-its-kind space-code beamforming technique to achieve ubiquitous radar coverage. We show that the range-angle map can be estimated for all targets at full range-resolution for all beams compared to existing subcarrier based beamforming radars.
\end{abstract}


\section{Introduction}
A truly ubiquitous radar differs from a mechanically rotating antenna based radar, and also from a conventional phased array radar in that it can "look everywhere simultaneously"\cite{s1_1}. Mechanically rotating antennas are limited by their fixed rotation rate, leading to fixed revisit time. Conventional phased array radar, while providing beam steering and variable revisit times, is limited by sequential beam processing and thus cannot achieve truly simultaneous coverage.

In \cite{s1_1}, Skolnik introduced the term ``ubiquitous phased array radar''. His vision was to enable a radar system to observe the entire surveillance volume and perform multiple functions concurrently, rather than sequentially.  The technical feasibility of realizing such a radar was explored in \cite{s1_2}, which also discussed implementation challenges and potential solutions. The approach in \cite{s1_2} employed a broad transmit beam with multiple narrow receive beams, formed using Digital Beamforming (DBF). While this wide transmit beam approach offers the advantages of improved coverage and implementation simplicity, it results in angular information being derived solely from the receive beam. This reliance on the receive beam for angle estimation can lead to potential error, particularly in near-far target scenarios. Specifically, a strong return from a near target received through the sidelobe may exceed the return from a far target received through the main lobe.  Such scenarios may result in significant leakage from sidelobe targets, and result in inaccurate target angle estimation. Several other researchers have adopted this wide transmit beam approach to enhance coverage, including the work reported in \cite{s1_4, s1_5, s1_7}.

To the best of our knowledge, published research has not addressed the problem of resource allocation for multiple beams, while also ensuring full range-resolution across all the beams. Since simultaneous multi-beam operation is our motive, time-domain beam switching is not considered. While frequency subcarriers can be allocated as a beam resource and split among multiple beams, as seen in subcarrier-based beamforming \cite{s1_8}, this would result in a reduction in range-resolution in each beam. Our work aims to overcome these limitations by achieving full range-resolution in all beams, while even attaining improved angular resolution through the use of multiple beams.

% The limitations of conventional radar systems, are exacerbated in scenarios requiring simultaneous long-range detection and short-range surveillance. The non-ideal solution of switching between these modes and processing them sequentially instead of simultaneous processing is a widely acknowledged challenge, especially in the military domain. This paper addresses this challenge by proposing an implementable system designed to achieve simultaneous multiple beams, hence simultaneous radar function processing, with full range-resolution in all beams. 

% An interesting implementation of Ubiquitous radar was shown in \cite{s1_5}, where the authors used conformal array instead of planar array. The paper discussed multiple configurations, ranging from low number of long range beams, to high number of short range beams. The conformal array is sectorised to do achieve these configurations. For the coexistence of one long range beam and multiple short range beams, the configuration was not discussed. We will explore such capabilities of coexistence of multiple different range beams, in this paper. 

\section{Proposed Signal Model}

The signal model is based on Zadoff-Chu (ZC) sequences \cite{s2_1}, which offer several desirable properties: unit amplitude, user-definable sequence length $N_{zc}$, the capability to generate multiple unique sequences, an ideal impulse-like periodic auto-correlation function (PACF), and fixed periodic cross-correlation function (PCCF) \cite{s2_2}. Furthermore, Fourier-duals of ZC sequences are also ZC sequences, retaining these properties under transformation \cite{s2_3}. ZC sequences are generated using \eqref{s1_zcsequence_gen} as shown below, parameterized by a seed value, $q$, and the length, $N_{zc}$, which are relatively prime to each other. 
A set of ZC sequences with length $N_{zc}$ is defined as $C = \{{Z_{q,N_{zc}}[n]}\:, 0 < n < N_{zc} - 1 \; | \; gcd(q, N_{zc}) = 1 \}$, where
\begin{equation}\label{s1_zcsequence_gen}
Z_{q,N_{zc}}[n] = 
\begin{cases} 
e^{j2\pi q{\frac{n^2}{N_{zc}}}} & \text{if } N_{zc} \text{ is even}, \\
e^{j2\pi q{\frac{n(n+1)}{N_{zc}}}} & \text{if } N_{zc} \text{ is odd} \\
\end{cases}
\end{equation}
As noted in \cite{s2_4}, sequences exhibiting ideal PACF are often employed in pulse compression radar due to their generally low aperiodic auto-correlation. We define periodic and aperiodic auto-correlation as follows,
\begin{equation} \label{s1_periodic_autocorr}
\begin{split}
R_{Z_{q, N_{zc}}}^{periodic}[j] = \sum_{n = 0}^{N_{zc}-1} Z_{q, N_{zc}}[(n+j)\bmod {N_{zc}}] Z^*_{q, N_{zc}}[n]
\end{split}
\end{equation}
\begin{equation} \label{s1_aperiodic_autocorr}
\begin{split}
R_{Z_{q, N_{zc}}}^{aperiodic}[j] = \sum_{n = 0}^{N_{zc}-1} \overline{Z}_{q, N_{zc}}[n+j] Z^*_{q, N_{zc}}[n] \\
\end{split}
\end{equation}
where $j = 0, 1, \dots , N_{zc} - 1$, and $Z^*_{q, N_{zc}}$ represents complex conjugate of  $Z_{q, N_{zc}}$. In \eqref{s1_aperiodic_autocorr}, $\overline{Z}_{q, N_{zc}}$ is $Z_{q, N_{zc}}$ appended with zeros before and after each of length $N_{zc}$, as expected from a pulsed radar receiver.

\begin{figure}[t]
\centerline{\includegraphics[width=0.45\textwidth, keepaspectratio]{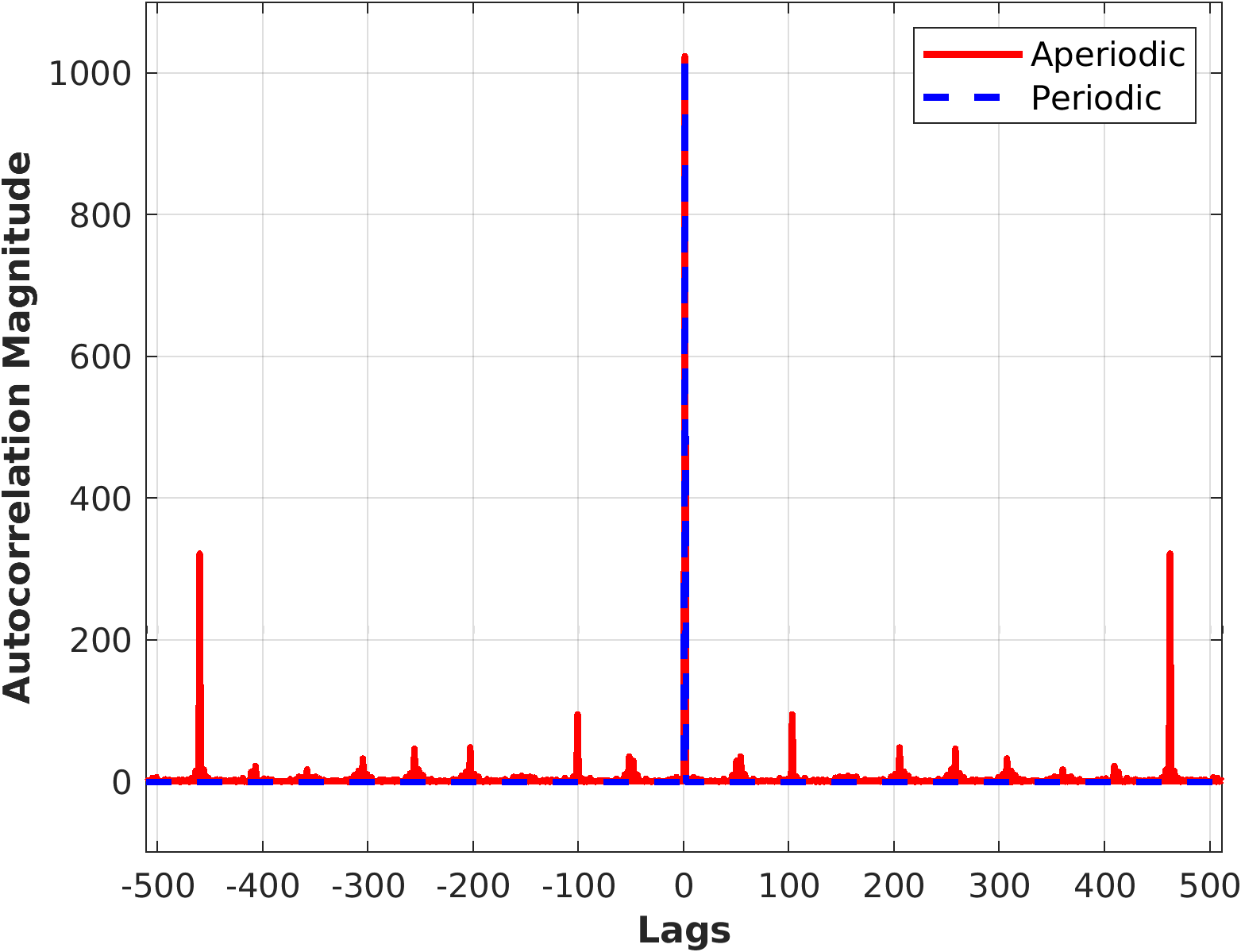}}
\caption{Comparison of periodic and aperiodic auto-correlation of 1024 length ZC sequence generated using seed 251. Note the presence of side peaks in the aperiodic auto-correlation, unlike the periodic auto-correlation.}
\label{s1_periodic_autocorr_img}
\vspace{-5pt}  % Reduces the space below the caption
\end{figure}

However, we observe that the perfect PACF of ZC sequence, as defined in  \eqref{s1_periodic_autocorr}, does not always guarantee low aperiodic auto-correlation, as defined in \eqref{s1_aperiodic_autocorr}. This is illustrated in Fig. \ref{s1_periodic_autocorr_img}, where the aperiodic auto-correlation function exhibits significant side peaks. These side peaks can introduce false alarms or mask weaker targets, thereby limiting the detector performance. In this section, we propose a signal model designed to ensure minimal aperiodic auto-correlation side-peaks.

\begin{figure}[b]
\centerline{\includegraphics[width=0.45\textwidth, keepaspectratio]{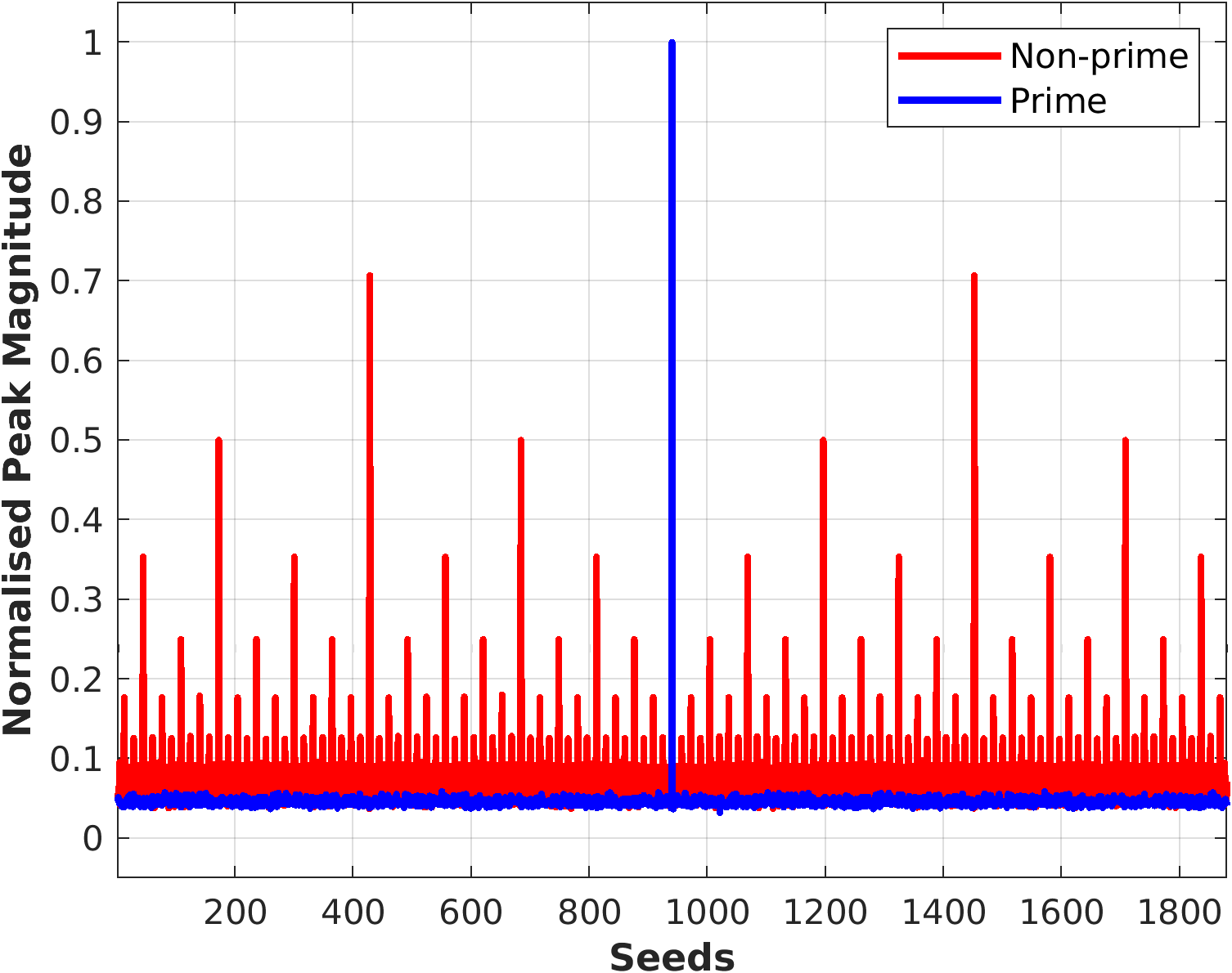}}
\caption{Normalised peak magnitude of cross-correlation between ZC sequence generated with seed 940 against all seeds, for 2 different lengths: Prime(1021) vs Non-prime(1024).}
\label{s1_prime_crosscorr_img}
\vspace{-5pt}  % Reduces the space below the caption
\end{figure}

We propose the use of multiple ZC sequence from the set $C$, keeping in mind that, minimising cross-correlation between these sequences is also vital. As established in \cite{s2_1}, selecting a seed $q$ and length $N_{zc}$ that are relatively prime yields perfect PACF. Furthermore, as described in \cite{s2_5}, PCCF is a function of $\sqrt{g.N_{zc}}$, where $g = gcd(N_{zc}, q_1 - q_2)$ and $q1$ and $q2$ are seeds of the cross-correlated sequence. It is clear that to achieve maximum PACF and minimum PCCF, two options exist: constrain the seed set to prime numbers with prime differences, or employ a prime-length ZC sequence. The latter approach ensures desirable PACF and PCCF properties are maintained regardless of seed selection. Figure \ref{s1_prime_crosscorr_img} shows that, unlike non-prime lengths, a prime $N_{zc}$ ensures minimum peak cross-correlation magnitude for all seeds. The sequence set is now modified to $C = \{Z_{q,N_{zc}}[n], 0 < n < N_{zc} - 1, \mid N_{zc} \in \mathbb{P}\}$, independent of seed $q$, where, $\mathbb{P}$ is the set of all prime numbers.

From the set $C$, ZC sequences with minimum side-peak in aperiodic auto-correlation must be selected to improve detection. To achieve this, we place the sequence $Z_{q,N_{zc}}$ in frequency domain as shown in \eqref{s1_sequenceplacement}, yielding the frequency-domain representation $S_q[k]$ as follows:
\begin{equation}
\begin{split}\label{s1_sequenceplacement}
S_q[k] = 
\begin{cases} 
0  & \text{, } k \ge\frac{-N}{2}+1 \text{ to }  k < \lfloor{\frac{-N_{zc}}{2}}\rfloor, \\ 
Z_{q,N_{zc}}[N_{zc} + k]& \text{, } k \ge \lfloor{\frac{-N_{zc}}{2}}\rfloor \text{ to } k < 0,\\
0  & \text{, } k = 0, \\ 
Z_{q,N_{zc}}[k-1]& \text{, } k > 0 \text{ to } k \le \lfloor{\frac{N_{zc}}{2}}\rfloor,\\
0 & \text{, } k > \lfloor{\frac{N_{zc}}{2}}\rfloor \text{ to } k \le \frac{N}{2}.
\end{cases} 
\end{split}
\end{equation}
where, $\lfloor\quad\rfloor$ denotes flooring to the previous integer and $N$ is a number greater than $N_{zc}$. The value of $N$ is even number and will be defined below.

In hardware implementation, we primarily use FFTs and IFFTs, which constraints the length of $S_q[k]$ to be a power of two. Consequently, the length of $S_q[k]$ is set to be $N=2^{\lceil log_2(N_{zc})\rceil}$, where $\lceil\quad\rceil$ denotes rounding to next integer. Due to Fourier duality property of ZC sequence, the correlation properties discussed above remain valid. This placement of prime-length ZC sequences onto the $N$ subcarriers as defined in \eqref{s1_sequenceplacement} also bandlimits the signal. Taking IFFT of $S_q[k]$ gives us time domain sequence $s_q[n]$, as follows:
\begin{equation}\label{s1_idft}
s_{q}[n] = \sum_{k = \frac{-N}{2} + 1}^{\frac{N}{2}} S_q[k] e^{j2\pi n\frac{k}{N}}, \quad n = 0,1 \dots N-1
\end{equation}

\begin{figure}[htbp]
\centerline{\includegraphics[width=0.45\textwidth, keepaspectratio]{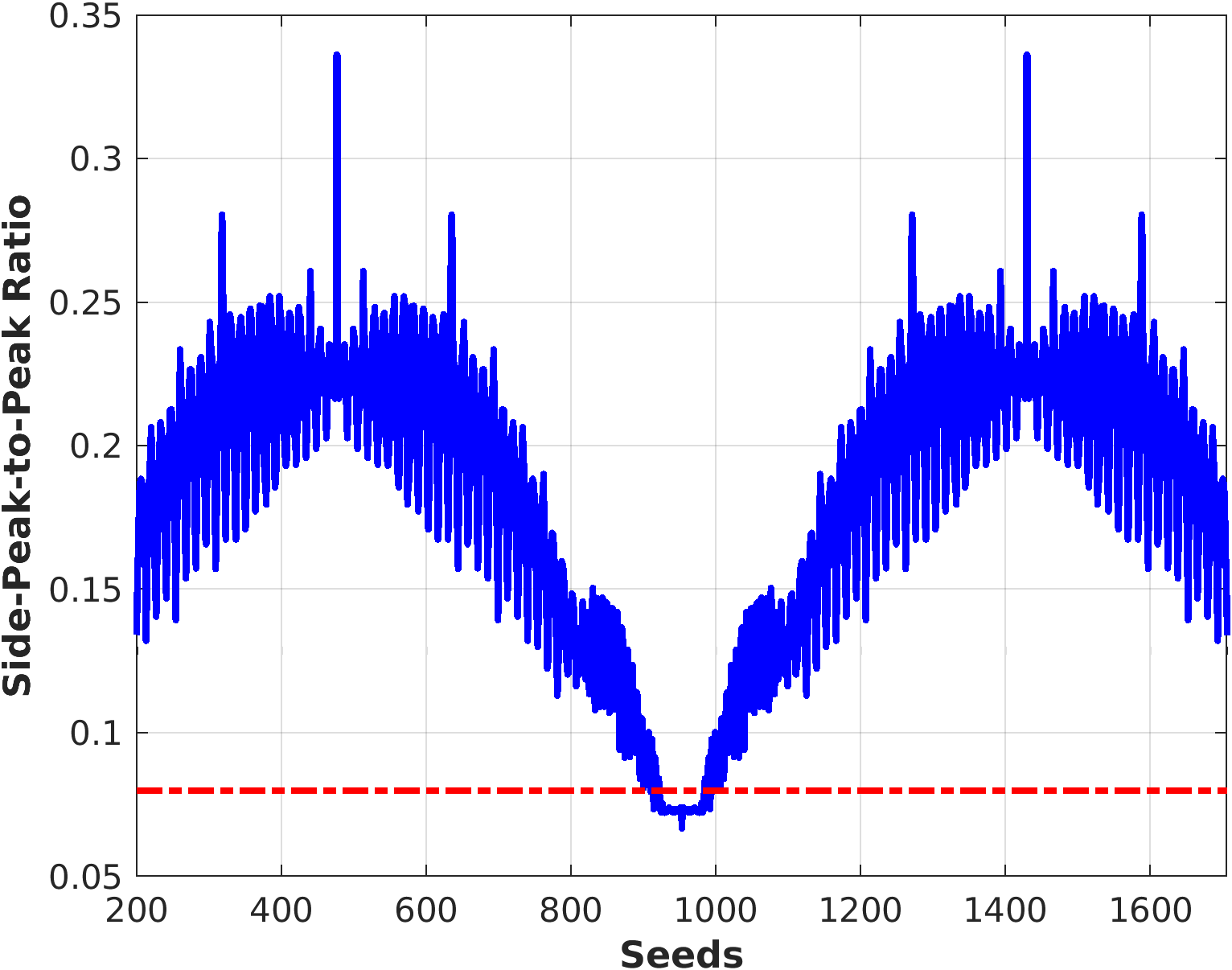}}
\caption{Side-Peak-to-Peak magnitude ratio is plotted for auto-correlation of prime-length (953) ZC sequence placed in 1024 subcarriers, for all seeds. The red dotted line puts a threshold, for values below this line the corresponding seeds produces minimum second peak in aperiodic auto-correlation, compared to remaining seeds.}
\label{s1_peak_to_secondpeak}
\vspace{-0pt}  % Reduces the space below the caption
\end{figure}

\begin{figure}[htbp]
\centerline{\includegraphics[width=0.45\textwidth, keepaspectratio]{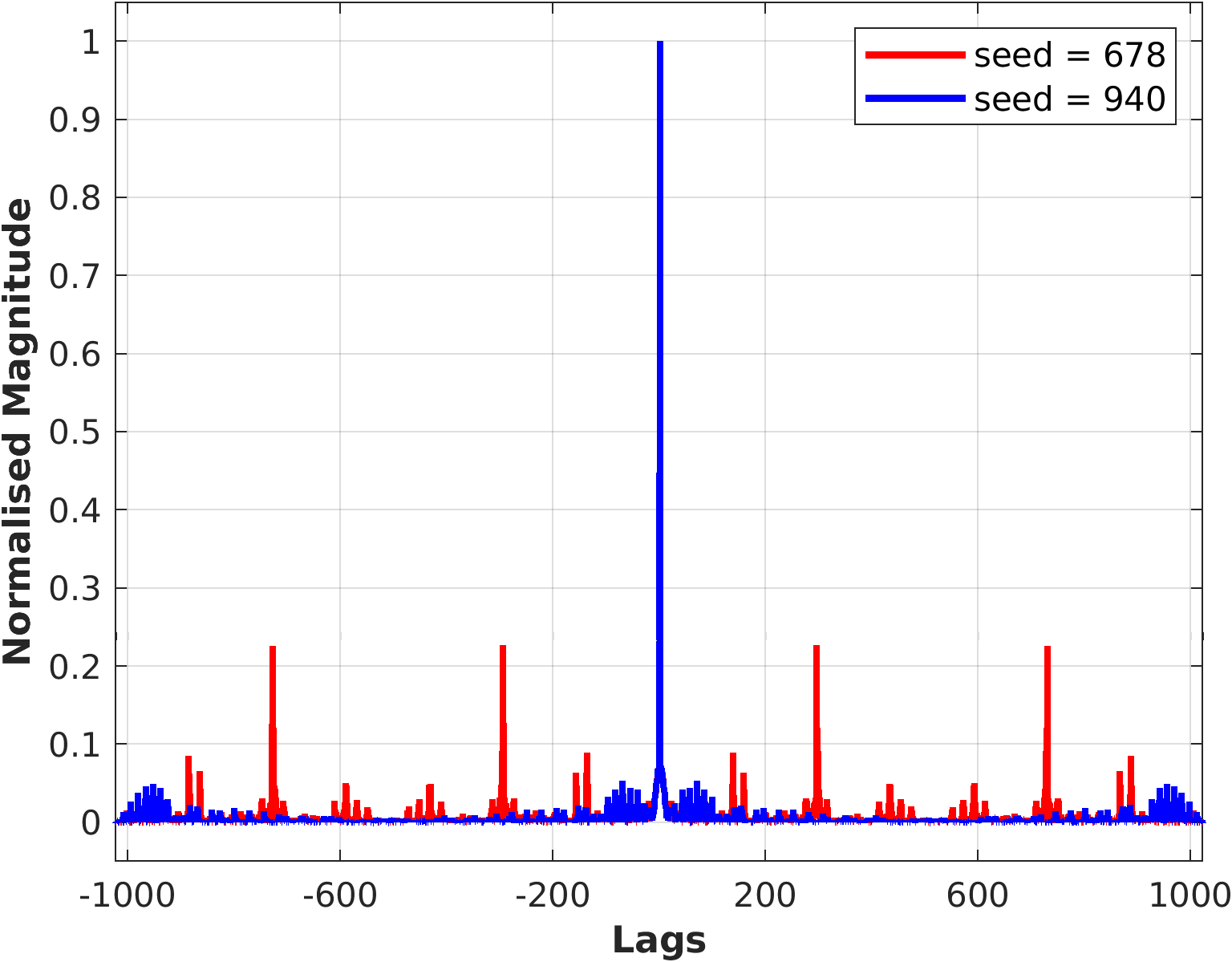}}
\caption{Normalised aperiodic auto-correlation magnitude of a prime-length (953) ZC sequence placed in 1024 subcarriers, for two seeds values: 940 and 678. The impact of seed selection on side peaks is displayed.} 
\label{s1_autocorr_freqdomain}
\end{figure}

Figure \ref{s1_peak_to_secondpeak}, shows the maximum side-peak-to-peak ratio of aperiodic auto-correlation of sequence $s_q[n]$. As the figure shows, seeds $q$ closer to the length $N_{zc}$ result in minimum achievable aperiodic auto-correlation, directly improving detector performance. The impact of seeds selection on aperiodic auto-correlation is plotted in Fig. \ref{s1_autocorr_freqdomain}.

This modifies our sequence set as $C = \{s_q[n], 0 < n < N - 1 \mid N_{zc} \in \mathbb{P},\; q \in [N_{zc} \pm B/2],\; q \neq N_{zc}\}$, where $B$ is the number of sequences needed and $\mathbb{P}$ is the set of all prime numbers. We conclude this section by noting that the generated sequence set $C$ exhibits impulse-like aperiodic auto-correlation and minimum cross-correlation among all ZC sequences. 

\section{Proposed Architecture}\label{architecture}

The success of ubiquitous radar heavily depends on the ability to use DBF to generate simultaneous multi-beam. In this section, we discuss the rationale behind choosing DBF over Analog Beamforming (ABF), followed by a description of transmit and receive chains.

A uniform linear array (ULA) is considered, consisting of $M$ transmit/receive (T/R) elements. The elements are spaced at half-wavelength $\frac{\lambda}{2}$ intervals to avoid grating lobes.  Element $m$, where $m \in [0, M-1]$, is located at a distance of $d_m = m\frac{\lambda}{2}$ from the zeroth reference element. The ULA is oriented horizontally, with its broadside facing the sky.

% In general DBF, the base-band chain generates $s_q[n]$, as shown in \eqref{s1_idft}, which is then sent to the delaying network. This network introduces delay $\tau_m$, given by \eqref{s2_delay}, and generates $M$ baseband signals $s_{q,m}[n - \tau_m f_s]$. Note that $\tau_m$ represents time in seconds, whereas $n$ is dimensionless sample number. Therefore $\tau_m$ is converted to samples by multiplying it with the sampling rate, $f_s$, resulting in $\tau_m f_s$. After digital-to-analog conversion, the signal $s_{q,m}(t-\tau_m)$ is generated and upconverted to passband, yeilding $Re\{s_{q,m}(t-\tau_m)e^{j2\pi f_c(t-\tau_m)}\}$. 
% \begin{equation} \label{s2_delay}
% \tau_{m} = \frac{d_{m}sin(\theta)}{c} \quad \text{where, } m = 0, 2, \dots, M-1
% \end{equation}

% Upon transmission from the M elements, the delays $\tau_m$ are perfectly compensated at an angle $\theta$ relative to the broadside direction. This results in an in-phase summation of the $M$ signals and consequently, an array gain equal to $M$. At angles other than $\theta$, the signals sums out of phase, leading to a reduction in power. This variation of array gain based on angle is captured as array pattern, it is given by \eqref{s2_array_pattern}
% \begin{equation} \label{s2_array_pattern}
% A(\phi) = \frac{1}{N}\sum_{k=-\frac{N}{2}+1}^{\frac{N}{2}} \sum_{m=1}^{M}e^{j2\pi (\tau_m - \frac{d_msin(\phi)}{c}) (f_c + \frac{f_sk}{N})} 
% \end{equation}
% where, $\phi \in (-180, 180)$, representing all possible azimuth angle.

Literature frequently cites the high cost and power consumption of DBF as disadvantages \cite{s3_1}. However, this assessment is applicable to scenarios where the DBF is employed to generate a single transmit beam. For our objective of generating multiple transmit beams, the relationship between cost and generating multiple beams must be understood. In DBF, delays are implemented as sample shifts which is computationally inexpensive. In contrast, ABF requires dedicated multiple hardware components for each delay element and hence for each beam too. Consequently, ABF systems exhibit a linear increase in hardware components, size, and cost with the number of beams. Furthermore, ABF offers limited flexibility, as the number of beams cannot be adapted or modified during runtime, a limitation not present in DBF systems.

As previously mentioned, implementing delays as sample shifts is computationally efficient. However, the product of $\tau_m f_s$ is generally not an integer, where $f_{s}$ is the sampling frequency and $\tau_m$ is the delay to be introduced, defined below in \eqref{s2_delay}. The delay value, $\tau_m$  depends beamforming angle $\theta$, the distance of the array element $d_m = m\frac{\lambda}{2}$ and speed of light $c$.
\begin{equation} \label{s2_delay}
\tau_{m} = \frac{d_{m}sin(\theta)}{c}
\end{equation}

This necessitates the introduction of arbitrary fractional delays to ensure accurate beamforming and avoiding undesirable array pattern distortion.  While time-domain interpolation techniques might seem feasible, they introduce significant processing overhead due to required upsampling and downsampling operations. Furthermore, the accuracy of the resulting delay is directly dependent on the interpolation factor, creating a trade-off between computational cost and precision. To mitigate this trade-off, frequency-domain beamforming \cite{s3_4} is employed. This approach enables the introduction of arbitrary delays through multiplication with a linear phase term in the frequency domain, as shown below.
\begin{equation}
\begin{split} \label{s2_freq_domain_beamforming}
S_q[k]  &\xrightarrow{\mathcal{F}^{-1}}  s_q[n] \\
S_q[k]e^{-j2\pi \tau_m f_s \frac{k}{N}} &\xrightarrow{\mathcal{F}^{-1}} s_q[n-\tau_m f_s]
\end{split}
\end{equation}

% \begin{equation} \label{s2_freq_domain_beamforming}
% \begin{split}
% s_q[n] &= \sum_{k=0}^{N-1}S_q[k] e^{j2\pi n\frac{k}{N}} \\
% s_q[n-\tau_mf_s] &= \sum_{k=0}^{N-1}S_q[k] e^{j2\pi n\frac{k}{N}} }
% \end{split}
% \xrightarrow{\mathscr{F^{-1}}}
% \end{equation}

\subsection{Transmit architecture}

In this system, we employ pulsed radar transceiver, where a sequence is transmitted, and the radar then switches to receive mode for a duration significantly longer than the transmission time. This approach eliminates the relationship between the maximum unambiguous range and the transmitted sequence length, a limitation inherent in continuous wave radars. Furthermore, the short transmit time given by $\frac{N}{f_s}$, simplifies the transmit chain and reduces power consumption. One complete cycle of transmission and reception is defined as the Pulse Repetition Interval (PRI).

The baseband transmit chain is shown below in Fig. \ref{s2_transmit}. The frequency domain signal $S_q[k]$ as defined in \eqref{s1_sequenceplacement}, is considered. To generate $B$ beams, $B$ distinct codes from set $C$ are employed, with each code mapped to a corresponding beam $S_q[k] \rightarrow S_b[k]$. Each code is directed towards a predefined space/angle $\theta_b$, where $b \in [0,B-1]$, via a dedicated delaying network. This means that to generate $B$ beams, we require $B$ delay networks. Every delay network produces $M$ outputs, sent to $M$ elements. Since we have $B$ such delay networks, each element receives $B$ signals which are summed. 

Note that all the processing till this point are done in the frequency domain, with the delay networks implemented by multiplying $S_b[k]$ with a linear phase term as shown in \eqref{s2_freq_domain_beamforming}. The resulting summed signal is then transformed into discrete time domain using an IFFT, yielding the Space-Code beamformed (SCB) signal for each element $m$, as:
\begin{equation}\label{s2_cdb_signal}
s_m[n] = \sum_{k = 0}^{N-1}\sum_{b = 0}^{B-1}\alpha_bS_b[k]e^{-j2\pi \tau_{m,b} f_s \frac{k}{N}} e^{j2\pi n\frac{k}{N}}
\end{equation}
where, $m = 0,\dots, M-1$, $n = 0,\dots, N-1$, $\alpha_b$ is power allocated to code $b$ and $\tau_{m,b}$ is given by,
\begin{equation} \label{s2__multi_delay}
\tau_{m, b} = \frac{d_{m}sin(\theta_b)}{c}
\end{equation}

\begin{figure}[htbp]
\centerline{\includegraphics[width=0.48\textwidth, keepaspectratio]{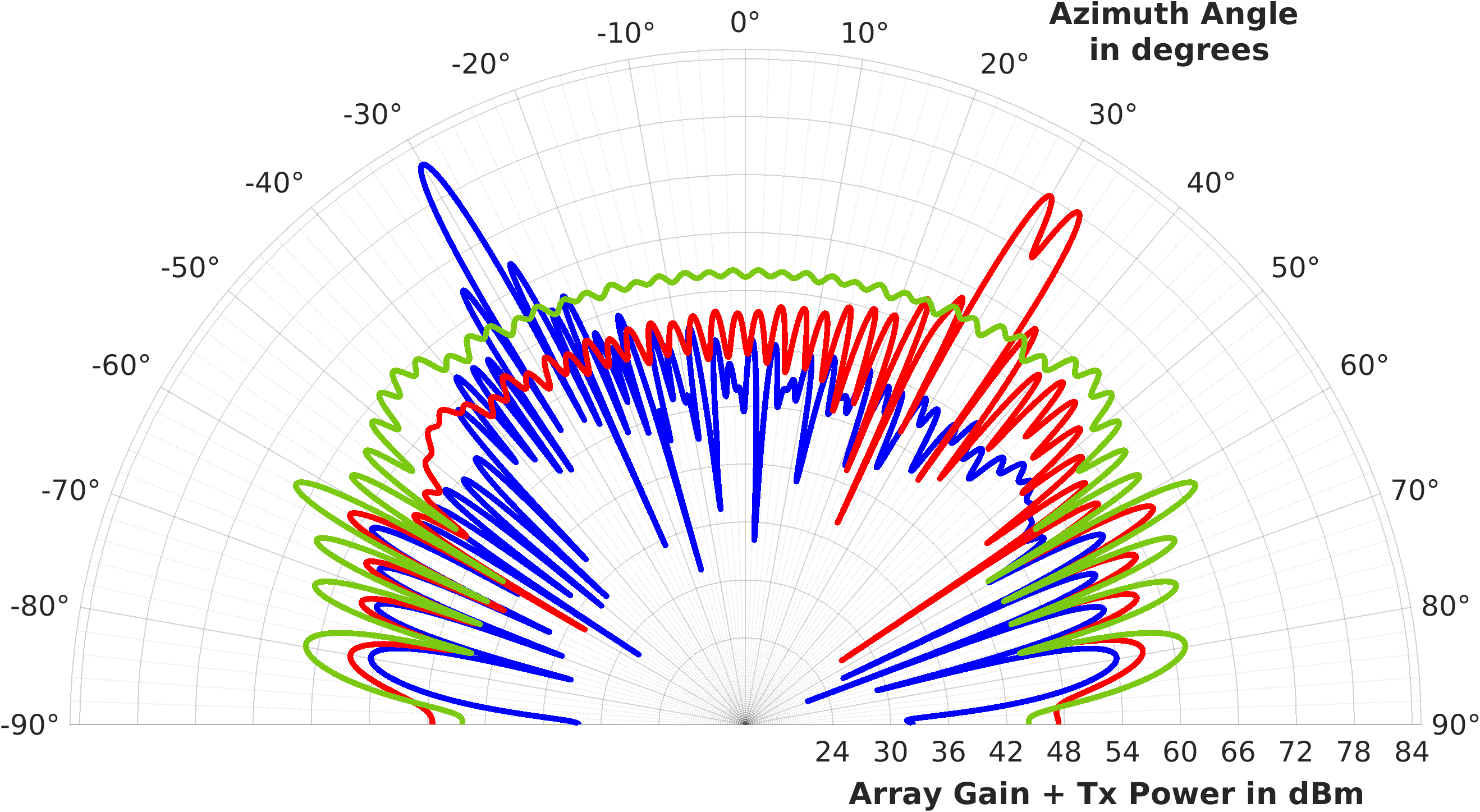}}
\caption{EIRP Pattern for 64-elements array, generating 41 beams from -60 to 60 deg at 3 deg spacing, with a constant power constraint (Total digital power = 42W) shown for three configurations: a) Blue - One Long range beam at -30 deg (39.204 W x 1 beam) + Reduced surveillance beam (0.064 W x 40 beams = 2.56 W), b) Red - Two Medium range beam at +30 and +33 deg (15.876 W x 2 beams) + Reduced surveillance beam (0.256 W x 39 beams = 9.984 W), and c) Green - 41 equal power surveillance beams (1.024 W x 41 beams = 41.984 W)} 
\label{s2_array_pattern}
\vspace{-15pt}  % Reduces the space below the caption
\end{figure}

In Fig. \ref{s2_transmit}, a power control block is incorporated to dynamically allocate the total dynamic range of the digital-to-analog converter (DAC) across different codes, and consequently, across the generated beams. This capability is demonstrated in Fig. \ref{s2_array_pattern}, showcasing ability to generate beams ranging from short-range, full-coverage surveillance beams to a few narrow, long-range beams. 
\vspace{-5pt}  % Reduces the space below the caption
\begin{figure}[htbp]
\centerline{\includegraphics[width=0.50\textwidth, keepaspectratio]{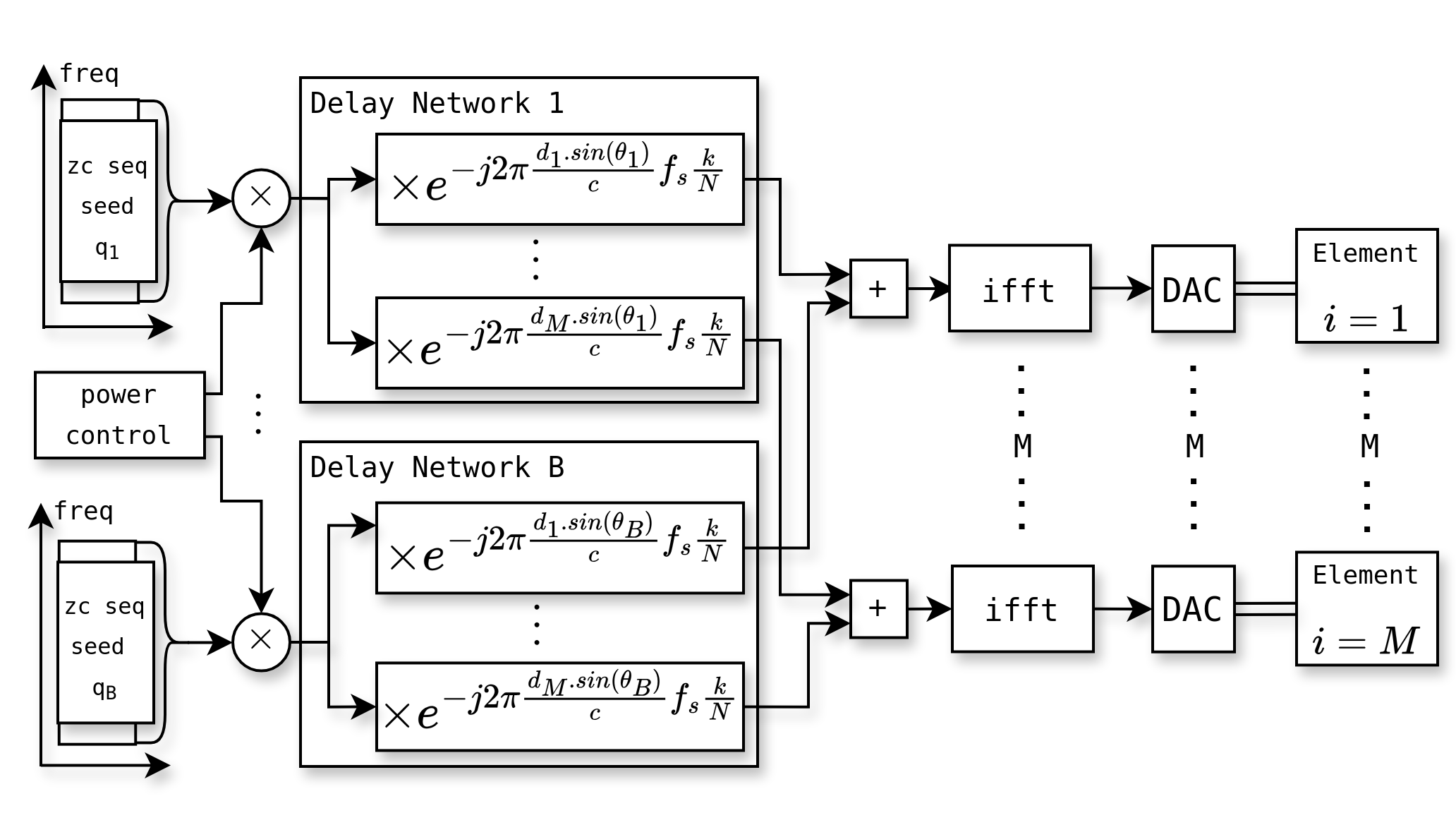}}
\caption{Space-Code Beamformer based Transmitter.} 
\label{s2_transmit}
\vspace{-0pt}  % Reduces the space below the caption
\end{figure}

The Effective Isotropic Radiated Power (EIRP) pattern, is the summation of array pattern and power transmitted in each code. We have plotted this EIRP pattern, instead of the array pattern, because SCB only manages the power transmitted in each code. Therefore, effective power variation over angles can be captured by the EIRP pattern, while the array pattern itself remains unchanged. Figure \ref{s2_array_pattern} highlights the angle-based power distribution flexibility afforded by the SCB technique.

\subsection{Receive architecture}

The receiver open time is determined by the maximum desired detection range of the system. For simplicity, we consider the receive open window to be k times the transmit time, $\frac{N}{f_s}$, which translates to a maximum detection distance of $k\frac{N}{2f_s}c$, where c is the speed of light. As described in the transmit chain subsection, multiple beams are transmitted simultaneously. These signals propagate through the medium, and if a target is present at a specific range-angle location, the reflection is returned towards the system and captured by the $M$ T/R elements in receive mode. For simplicity, the target's Radar Cross Section (RCS) is set to 0 dBsm, indicating that the target reflects all incident power.

The signals received from the $M$ elements must be digitally beamformed to generate multiple receive beams. A key challenge is that the received signal duration is $k$ times longer than the transmitted signal, resulting in a continuous data stream. This contrasts with the transmitted signal, where beamforming was performed on a data block, rather than a data stream.

\begin{figure*}[t]
\centerline{\includegraphics[width=\linewidth, keepaspectratio]{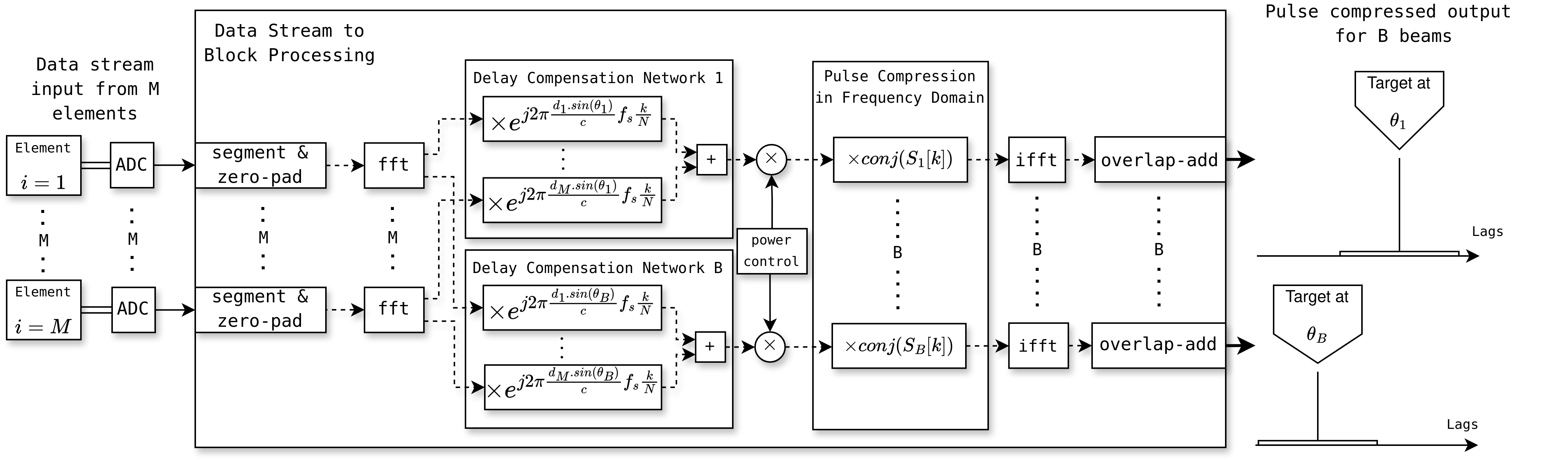}}
\caption{Space-Code Beamformer based Receiver with parallel processing of B beams, and pipelined with pulse compression performed in frequency domain.} 
\label{s2_code_domain_bf_rx}
\vspace{-0pt}
\end{figure*}

To address this, we propose a receiver chain that combines the well-known overlap-add method \cite{s3_5} with SCB to process the data stream received. The receive chain is illustrated in Fig. \ref{s2_code_domain_bf_rx}. The overlap-add method emulates block-wise processing on a continuous data stream. This is achieved by sequentially taking $N$ samples from the data stream, padding each segment with $N$ zeros and then perform a $2N$-length FFT. The resulting $2N$-length data blocks from $M$ elements are processed by $B$ individual delay compensation networks. Each delay compensation network corrects the delay by multiplying the frequency domain signal with a linear phase term and summing the $M$ signals to produce a single output per network, resulting in $B$ total outputs, for $B$ networks.

Leveraging the frequency domain, we also perform simultaneous pulse compression on $B$ signals corresponding to each beam by multiplying with complex conjugate of twice oversampled $S_b[k]$, due to $2N$-length, which is precalculated and stored. This approach is computationally more efficient than direct-form convolution of an $N$ length sequence, especially as $N$ increases \cite{s3_7}. Having processed the data blocks entirely in frequency domain, we now perform a $2N$-length IFFT on all blocks and complete the overlap-add process by adding the blocks with an overlap of $N$ samples from the preceding block, effectively converting the block-based processing back into a continuous output stream. 

It is important to note that all these processes occur in parallel for all beams/angles, with beamforming and pulse compression performed in a pipelined fashion, enabling the system to ``look everywhere simultaneously''. These $B$ outputs can then be directly passed to the detector.

% since we do, receive beamforming, we can see the main lobe has twice the power gain from transmit and receive and code gain also.

% table to compute number of ffts, iffts and multiplication for transmit and receive chain
% \renewcommand{\arraystretch}{1.5} 
% \begin{table}[htbp]
% \centering
% \caption{Implementation Complexity}
% Using the "tabular" environment with customized column spacing
% \begin{tabular}{|p{3cm}|p{2cm}|p{2cm}|}
% Top border line (using booktabs for cleaner lines)
% \hline 
% \textbf{Parameter} & \textbf{No. of FFTs} & \textbf{No. of IFFTs}   \\ 
% \hline
% Transmit chain & - & $M$\\
% \hline
% Receive chain & $M$ & $B$\\
% \hline
% \end{tabular}
% \vspace{7pt}
% \label{implementation_complexity}
% \end{table}
% \vspace{-5pt} 

\section{Simulation Results}
A simulation setup is considered, employing the $M$-elements ULA described in Section \ref{architecture} and the parameters specified in the table \ref{simulation_parameters}, with an equal-power EIRP pattern as illustrated in Fig. \ref{s2_array_pattern}c.

\vspace{-5pt} 
\renewcommand{\arraystretch}{1.5} 
\begin{table}[htbp]
\centering
\caption{Simulation Parameters}
% Using the "tabular" environment with customized column spacing
\begin{tabular}{|p{4cm}|p{2.2cm}|}
% Top border line (using booktabs for cleaner lines)
\hline 
\textbf{Parameter} & \textbf{Value}  \\ 
\hline
Carrier Frequency, $f_c$ & 10 GHz\\
\hline
Sampling Frequency, $f_s$ & 10 MHz\\
\hline
Sequence length, $N$ & 2048\\
\hline
ZC Sequence length, $N_{zc}$ & 1931\\
\hline
Bandwidth, $\frac{f_sN_{zc}}{N}$ &  9.42 MHz\\
\hline
No. of T/R elements, $M$ & 64 elements \\
\hline
No. of Beams, $B$ & 41 beams\\
\hline
Angular spacing, $\Delta\theta$ & $3^\circ$ spacing\\
\hline
Azimuth angles, $\theta_B$ & $-60^\circ: \Delta\theta :60^\circ$\\
\hline
Amplitude per code/beam, $\alpha_b$ & 32 \\
\hline
Total Digital Power, $\alpha_b^2B$ & 41984 mW $\simeq$ 42W\\
\hline
Target RCS & 0 dBsm\\
\hline
\end{tabular}
\vspace{0pt}
\label{simulation_parameters}
\end{table}
% \vspace{pt} 

\begin{figure}[htbp]
\centerline{\includegraphics[width=0.50\textwidth, keepaspectratio]{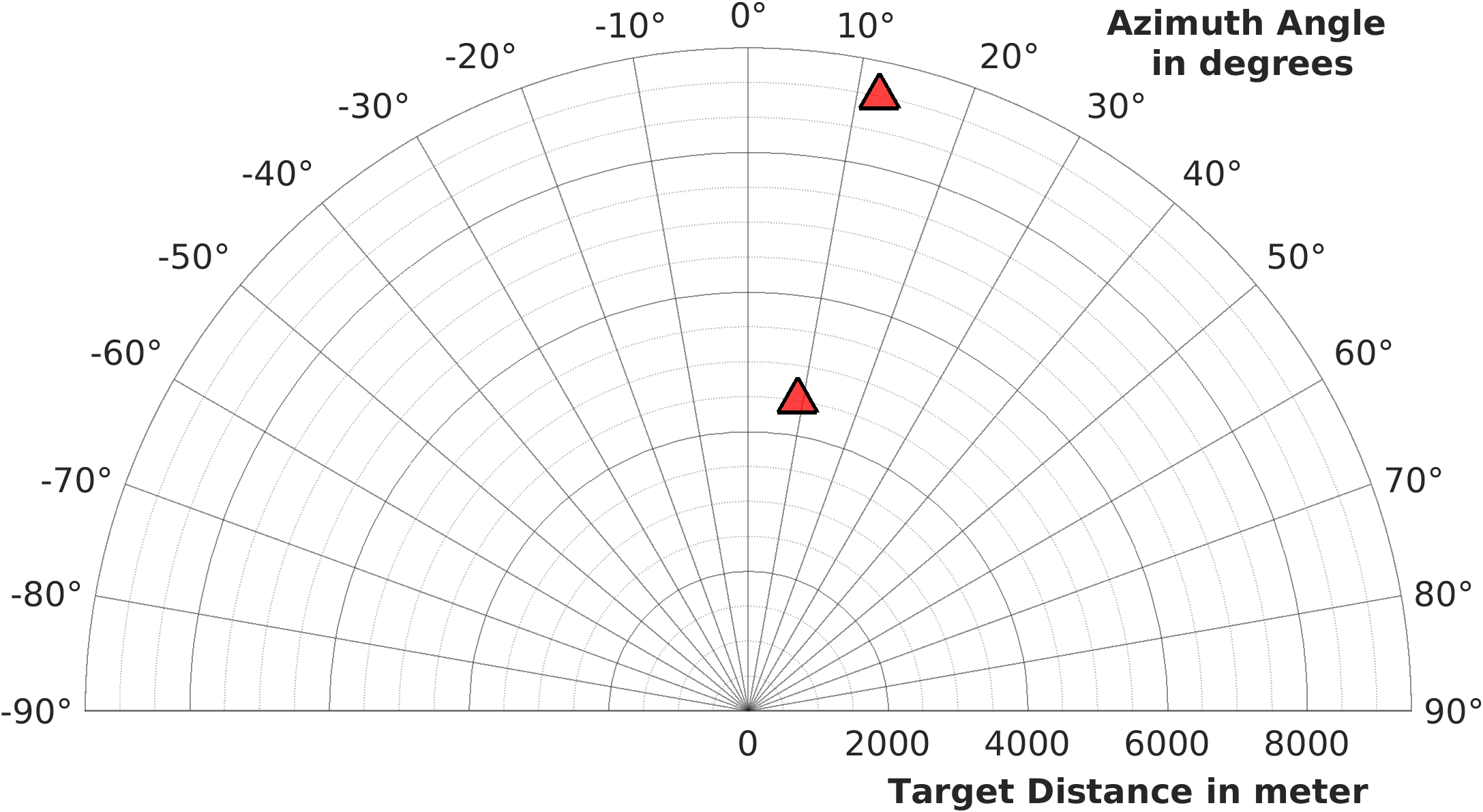}}
\caption{Two targets placed at [9 deg, 4.5 Km] and [12 deg, 9 Km]. Both targets are in adjacent beams formed at 9 deg and 12 deg.} 
\label{s3_target_map1}
\vspace{-10pt} 
\end{figure}

\begin{figure}[htbp]
\centerline{\includegraphics[width=0.50\textwidth, keepaspectratio]{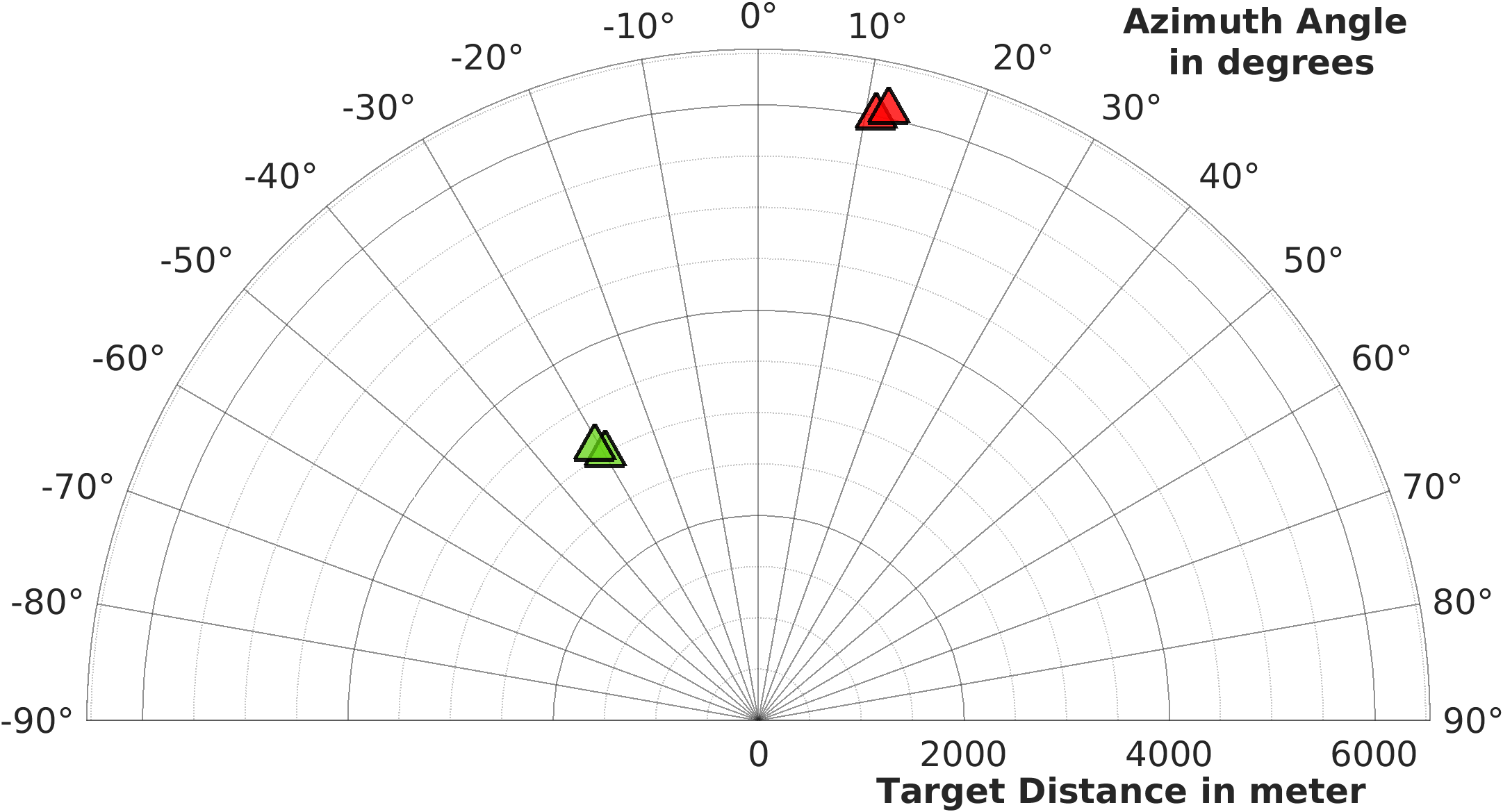}}
\caption{Four targets are paired as two sets of closely spaced targets in different beams. Set 1 (Green) is placed at 3000 meters and 3070 meters, at -30 deg beam. Set 2 (Red) is placed at 6000 meters and 6070 meters, at 12 deg beam.}
\label{s3_target_map2}
\vspace{-5pt}  % Reduces the space below the caption
\end{figure}

To demonstrate the advantages of SCB technique, we consider two experiments: two targets positioned in adjacent beams with a large range difference as shown in Fig.\ref{s3_target_map1}, and four targets, arranged as two pairs of closely spaced targets, positioned in two different beams as shown in Fig.\ref{s3_target_map2}. 

\subsection{Single Code vs Multi Code}

This section compares multiple codes mapped to multiple beams to using a single code for all beams.

% \vspace{-10pt} 
\begin{figure}[htbp]
\centerline{\includegraphics[width=0.50\textwidth, keepaspectratio]{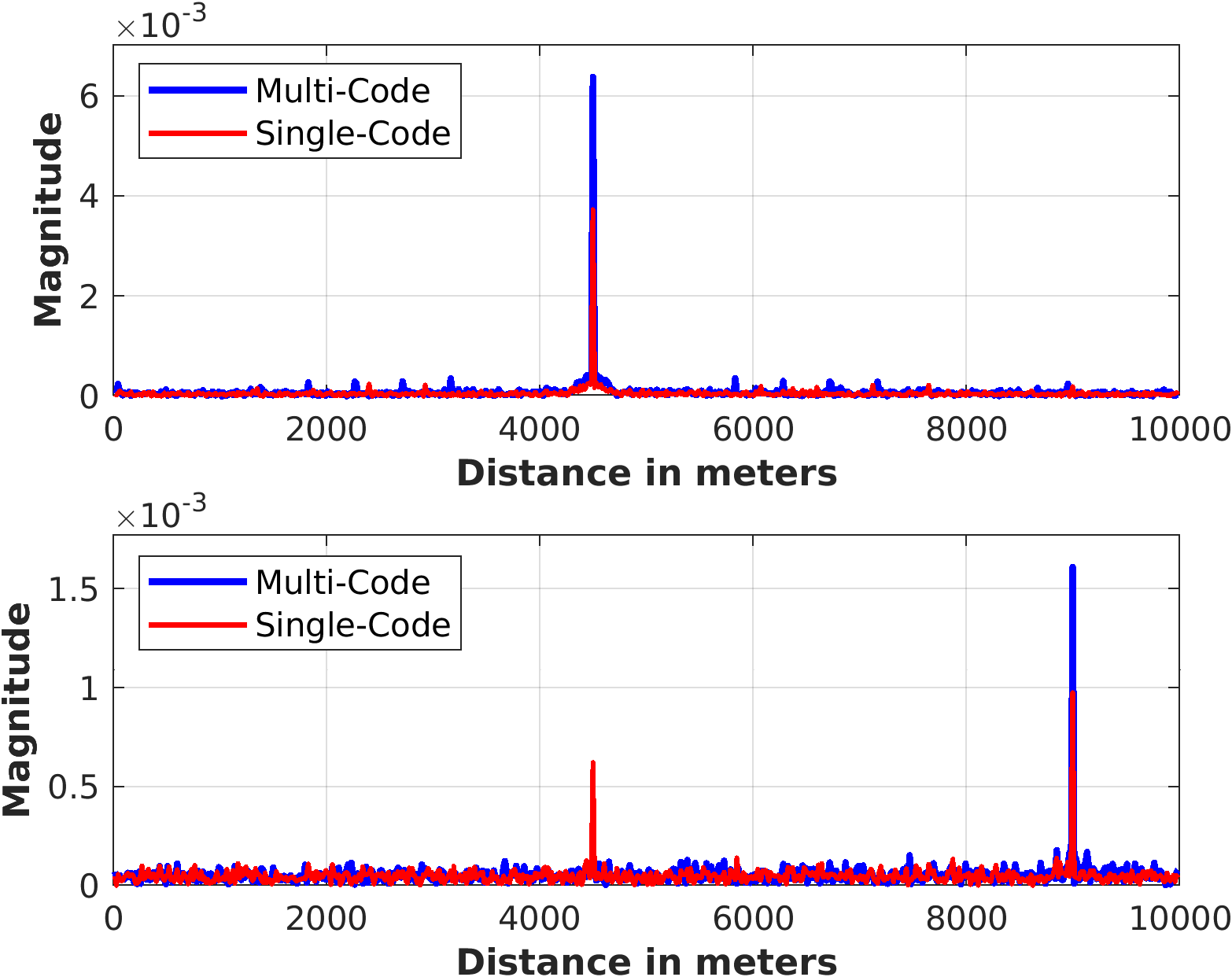}}
\caption{Comparison of pulse compression output (top: 9 deg beam, bottom: 12 deg beam). Multi-code effectively mitigates target leakage from the 9 deg beam to the 12 deg beam, unlike the single-code.}
\label{s3_angular_res}
\vspace{-10pt}
\end{figure}

Figure \ref{s3_angular_res} illustrates that in the single-code output, the target from the $9^\circ$ beam (top) appears as a false target in the $12^\circ$ beam (bottom) pulse-compressed output due to sidelobe leakage. In contrast, the multi-code pulse-compressed output shows no ghost target from adjacent beam. Despite the closer proximity and lower propagation loss of the target at $9^\circ$ beam, it is not present in the $12^\circ$ beam multi-code output, demonstrating the superior angular distinction of the multi-code approach.

In other words angular estimation is enhanced by utilizing code space in conjunction with beam space. This enhancement arises because SCB leverages information from both receive beams and the transmit beam associated with each code.

\subsection{Space-Code vs Space-Subcarrier mapped beamforming}
This section highlights the capability of this system to provide full range-resolution i.e., ability to distinguish closely spaced targets for all beams simultaneously.

\begin{figure}[htbp]
\centerline{\includegraphics[width=0.50\textwidth, keepaspectratio]{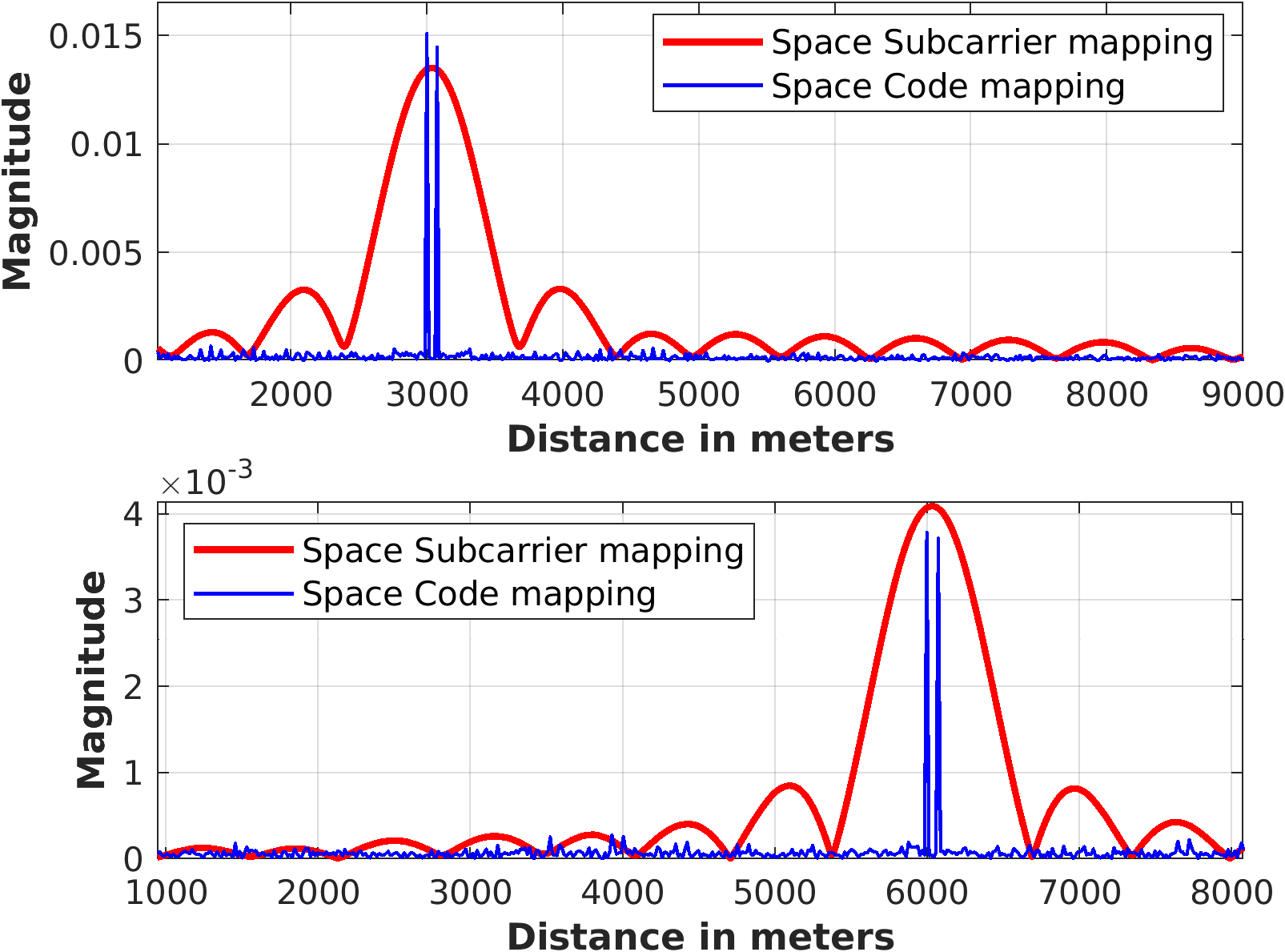}}
\caption{Pulse compression output (top: -30 deg, bottom: 12 deg beams). Space code mapping differentiates two close targets using full range-resolution, but subcarrier mapped beamforming loses resolution and merges both targets as single target. Space subcarrier mapping output is scaled for display.} 
\label{s3_range_res}
\vspace{0pt}  % Reduces the space below the caption
\end{figure}

From Fig. \ref{s3_range_res}, it is evident that, SCB achieves full band allocation across all beams, maintaining the capability to resolve closely spaced targets and ensuring full range-resolution in all beams. Conversely, subcarrier-based allocation leads to a merging of closely spaced targets, which significantly limits the potential performance of the system.

\section{Conclusion}

The ability to achieve full range-resolution in all beams simultaneously through the use of a novel Space-Code Beamforming, is a key contribution of this paper. This, combined with improved angular estimation and all-frequency domain pipelined implementation, represents a significant advancement for future radar systems.  Furthermore, the flexibility to perform both narrow-beam based far-target localization and wide-area near-field surveillance in a single system, while processing all beams simultaneously, represents a significant step towards realizing the vision of ubiquitous radar.

\vspace{12pt}

\end{document}